\documentclass[%
reprint,
superscriptaddress,
twocolumn,
 amsmath,amssymb,
 aps,
 prx,
]{revtex4-1}

\usepackage{graphicx}% Include figure files
\usepackage{dcolumn}% Align table columns on decimal point
\usepackage{bm}% bold math
\usepackage{amsmath}  % improve math presentation
\usepackage{amssymb}
\usepackage{xcolor}
\usepackage{braket}
\begin{document}

\preprint{APS/123-QED}

\title{Identifying f-electron symmetries of UTe$_2$ with O-edge resonant inelastic X-ray scattering}% Force line breaks with \\
% \title{Valence and symmetry in UTe$_2$ from O-edge resonant inelastic X-ray scattering}

\author{Shouzheng Liu}
\author{Yishuai Xu}
\author{Erica C. Kotta}
\affiliation{Department of Physics, New York University, New York, New York 10003, USA}
\author{Lin Miao}
\affiliation{School of Physics, Southeast University, Nanjing 211189, China.}
\author{Sheng Ran}
\affiliation{Department of Physics, Washington University in St. Louis, St. Louis, MO 63130, USA}
\author{Johnpierre Paglione}
\affiliation{Quantum Materials Center, Department of Physics, University of Maryland, College Park, MD 20742, USA}
\author{Nicholas P. Butch}
\affiliation{NIST Center for Neutron Research, National Institute of Standards and Technology, 100 Bureau Drive, Gaithersburg, MD 20899, USA}
\affiliation{Quantum Materials Center, Department of Physics, University of Maryland, College Park, MD 20742, USA}
\author{Jonathan D. Denlinger}
\author{Yi-De Chuang}
\affiliation{Advanced Light Source, Lawrence Berkeley National Laboratory, Berkeley, CA 94720, USA}
\author{L. Andrew Wray}
\email{lawray@nyu.edu}
\affiliation{Department of Physics, New York University, New York, New York 10003, USA}

% \affiliation{Physics Department, New York University.}%Lines break automatically or can be forced with \\
%\affiliation{%


\date{\today}

\begin{abstract}
The recent discovery of spin-triplet superconductivity emerging from a non-magnetic parent state in UTe$_2$ has stimulated great interest in the underlying mechanism of Cooper pairing. Experimental characterization of short-range electronic and magnetic correlations is vital to understanding these phenomena. Here we use resonant inelastic X-ray scattering (RIXS), X-ray absorption spectroscopy (XAS), and atomic multiplet-based modeling to shed light on the active debate between 5$f^2$6$d^1$-based models with singlet crystal field states versus 5$f^3$-based models that predict atomic Kramers doublets and much greater 5$f$ itinerancy. The XAS and RIXS data are found to agree strongly with predictions for an 5$f^2$6$d^1$-like valence electron configuration with weak intra-dimer magnetic correlations, and provide new context for interpreting recent investigations of the electronic structure and superconducting pairing mechanism.
\end{abstract}

\maketitle

The compound UTe$_2$ has been subject to recent attention following the discovery that it hosts spin-triplet superconductivity ($T_{c}\sim1.6K$) emerging from a nonmagnetic parent state, resembling a solid state analogue of superfluid He$^3$ \cite{ranshen_early_paper,Ute2_upper_critical_field,Ute2_TRSB}. This exciting discovery suggests that the material may host Majorana boundary modes of interest for fault-tolerant quantum computation \cite{majorana}. Theoretical modeling has predicted a strong ferromagnetic (FM) interaction between dimerized uranium atoms, which is widely speculated to be a driver of the triplet pairing \cite{yang_paper,qimiao_paper,anderson_model}. However, experimental investigations of the uranium 5$f$ electron configuration have been subject to divergent interpretations \cite{fujimori_paper4,miao_paper,charge_density_resolved_paper,UTe2_dHvA}. Here we combine evidence from RIXS, XAS, and atomic multiplet-resolving modeling to assess the multiplet symmetry and intra-dimer interactions of uranium electrons, showing that a fully consistent picture is achieved with an effective 5$f^2$6$d^1$ valence state.

% paragraph 1.5: add any structure/etc details needed to understand our results - maybe the charge density resolved measurement results?
The UTe$_2$ lattice is orthorhombic and belongs to the Immm space group \cite{Ute2_structure}. Two uranium atoms in each unit cell form a closely spaced dimer structure separated along the c-axis, and these dimers act as the rungs of a quasi-one dimensional ladder that runs along the a-axis. Unlike other spin-triplet candidates such as UGe$_2$ \cite{UGe2_paper}, URhGe \cite{URhGe_paper}, and UCoGe \cite{UCoGe_paper} in the family of uranium-based superconductors, no long range magnetic order is observed at ambient pressure for UTe$_2$ \cite{ranshen_early_paper}. Momentum resolved inelastic neutron scattering (INS) experiments have observed antiferromagnetic fluctuations at the (0,0.57,0) wavevector within UTe$_2$ \cite{INS_2,Butch_neutron}, and an inelastic structure factor consistent with intradimer FM correlations \cite{INS_1}. However, such scattering features have not been observed in the elastic channel, and require further research to fully integrate in a broader picture of the electronic structure. Core level measurements of uranium charge density are suggestive of a U$^{3+}$-like or intermediate U$^{3.x+}$ valence state \cite{charge_density_resolved_paper}, a result that has been separately interpreted as supporting both 5$f^2$6$d^1$ and 5$f^3$ effective valence pictures \cite{miao_paper,fujimori_paper4}.
%we can remove the reference to INS_2 if we need space
%The researchers reported that they discovered the dominant fluctuations near the AF vectors (0,0.57,0) by aligning 61 UTe$_2$ samples of average mosaic ~15°.

%2nd paragraph: experimental details (look at 2nd paragraph of URu2Si2 2015 PRL RIXS paper; note that the penetration depth on resonance is shallow (effectively several nanometers ***I can add citations), but is in principle large relative to techniques such as ARPES/STM
Measurements at the uranium O-edge were performed under ultra-high vacuum ($P<4\times 10^{-10}$ Torr) at the ALS BL4.0.3 MERIXS endstation, with better than $\delta E<50$ meV RIXS resolution at $h\nu = 100$ eV. Large $\sim$1 mm$^3$ samples were cleaved \emph{in situ} at T=20K along the [011] surface, and aligned to include the [100] axis within the scattering plane. A near-normal 23$^\circ$ angle of beam incidence was used for all measurements, and RIXS measurements were performed with $\pi$ polarization and a 90$^\circ$ scattering angle to reduce the intensity of elastic scattering. The penetration depth of both XAS and RIXS was comparable to $\gtrsim 2$ nm, and RIXS provides larger penetration depths throughout most of the spectrum. This is in principle large enough to sample bulk-like properties but does not rule out surface-derived spectral features.

Cross sections for RIXS and XAS were obtained from atomic multiplet modeling. Slater-Condon terms were obtained from first principles Hartree-Fock calculations, with renormalization of 70$\%$ for 5$f$-5$f$ interactions, 60$\%$ for 5$d$-5$f$ interactions, and 85$\%$ for 5$f$ spin orbit coupling, comparable to Ref. \cite{wray2015}. Calculations are Boltzmann weighted to temperatures indicated in the text. Magnetic coupling within a uranium dimer is assigned a J=50 meV ferromagnetic exchange constant \cite{yang_paper}. As only one uranium atom is explicitly included in the modeling basis and the crystal field is not fully understood, the amplitude of the intradimer exchange perturbation is obtained from the expectation value of easy-axis moment amplitude versus temperature on the neighboring site in an Ising-like Boltzmann weighted 2-atom model with no crystal field. This is expected to provide an overestimate (or upper bound) of the easy-axis magnetic exchange perturbation. The approximate crystal field symmetry is obtained by acting on uranium orbitals delta function potentials at the 8 nearest-neighbor Te coordinates identified in \cite{simu_ref1}. The 6 further Te neighbors are taken to apply identical perturbations on $z^3$ orbitals aligned with the relevant U-Te axis, while the closer 2 atoms are assigned a 25\% larger perturbation. This symmetry reproduces the easy and hard axes ($a$ and $b$ respectively) for magnetic polarization. The amplitude of the crystal field is set to give singlet states at energies corresponding to $E/k_B$= 0K, 40K and 139K, consistent with \cite{simu_ref2}.

\begin{figure}[ht]
    \centering
    \includegraphics[width=0.5\textwidth]{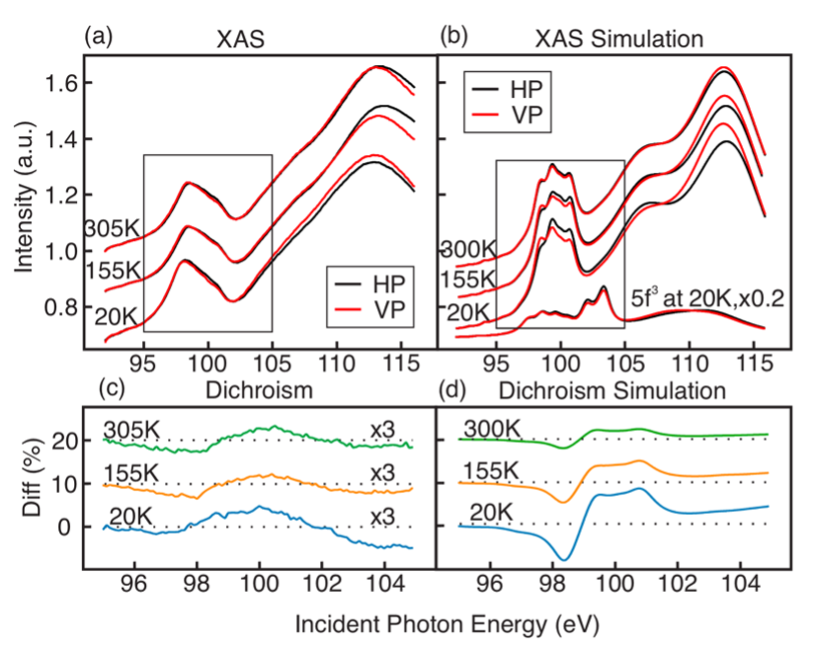}
    \caption{Atomic f-electron symmetries from X-ray absorption. (a) XAS curves for UTe$_2$ at different temperatures. Horizontal (HP, black) and vertical (VP, red) incident photon polarization is indicated by the color of the curve. (b) Multiplets simulations for 5$f^2$ and 5$f^3$ configurations. The 5$f^2$ simulation is shown for the same temperatures as the data in panel (a). (c,d) The dichroic difference (HP-VP) of the XAS scan and multiplet simulations, normalized by the feature height of the $h\nu\sim99$ eV resonance. Curves in panel (c) are enlarged by a factor of 3 for better visibility.}
    \label{fig:fig1}
\end{figure}

% 3rd paragraph: Segue sentence: explain that the O-edge can be used to resolve orbital symmetry, and so we begin with XAS (Fig. 1). then Describe Fig. 1 
Uranium O-edge resonant spectroscopies have recently been identified as powerful tools for identifying the multiplet symmetry of uranium 5$f$-electrons, which is generally not resolved at other resonance edges \cite{O_edge_importance,wray2015,miao_USb2,amorese_paper}. Curves in Fig. \ref{fig:fig1}(a-b) show XAS measurements on a UTe$_2$ sample at three different temperatures alongside 5$f^2$ atomic multiplets simulations. A 5$f^3$ simulation is also shown in Fig. \ref{fig:fig1}(b), and reveals a very different spectrum with a more equal branching ratio between the low and high energy resonances at $h\nu\sim$99 and $\sim$110 eV as well as a prominent resonance at $h\nu\sim$103 eV that is not visible in the experimental data. The enhanced leading edge of the experiment relative to the simulation is a common feature with earlier measurements on 5$f^2$ systems \cite{wray2015}, and may relate to the trend toward greater energy-axis broadening of excitations at higher incident photon energy within a resonance \cite{Kotani_lifetime,Sawatzky_lifetime,Frontiers_RIXS_review}.

% %if needed, add 1 paragraph discussing implications and physics
Linear dichroism of the XAS curves is presented in Fig. \ref{fig:fig1}(c) to more closely investigate the ground state symmetry and interplay with magnetic correlations. The lineshape of dichroism features a dip at the leading edge ($\sim$98 eV) followed by a region of positive intensity on top of the $\sim$100 eV resonance. Higher energy features are not closely analyzed as they are susceptible to strong bias from Fano interference in the photoemission process \cite{wray2015}. The sign and global amplitude of X-ray linear dichroism is determined by the multiplet symmetry of f-electrons \cite{linear_dichroism_quadrupole}, however the lineshape is fixed for a 5$f^2$ multiplet simulation (see Fig. \ref{fig:fig1}(c)) and bears a dip and peak that correspond to the experiment. Our model yields a matching global sign for the dichroism curves, which supports the accuracy of the simulated crystal field. However,the experimental amplitude is smaller by a factor of $\gtrsim 3$ at low temperature, and does not appear to evolve with temperature as would be expected for a system governed by single-atom 5$f^2$ multiplet physics \cite{miao_paper} (see Fig. \ref{fig:fig1}(d) simulation).

%However, the amplitude and sign of these features depend on the degree of alignment of total angular momentum with an axis defined by the scattering geometry. Our modeling shows consistency with experiment, but the experimental amplitude is small and it is not clear if the amplitude changes with temperature as it would be expected to do in a system with highly localized 5$f^2$ multiplet physics \cite{miao_paper}. 
%*** we need to explain later that scattering with the 6d electrons can explain the small amplitude and lack of temperature dependence; but that the system is also near the dichroic sign reversal point, so one should not read too much into this

One possible explanation for the discrepancy is that scattering that exchanges angular momentum between 5$f$ and 6$d$ electrons will result in an ensemble of 5$f$ multiplet symmetries, which on average is expected to suppress the amplitude of the dichroism curves. One should also note that the 5$f^2$ crystal field ground state can undergo dichroic sign reversal with only rather minor tuning of the crystal field parameters. Hence the small experimental dichroic amplitude could potentially be accounted for through fine tuning of the crystal field, and does not necessarily require additional physics.

\begin{figure*}
    \includegraphics[width=0.8\textwidth]{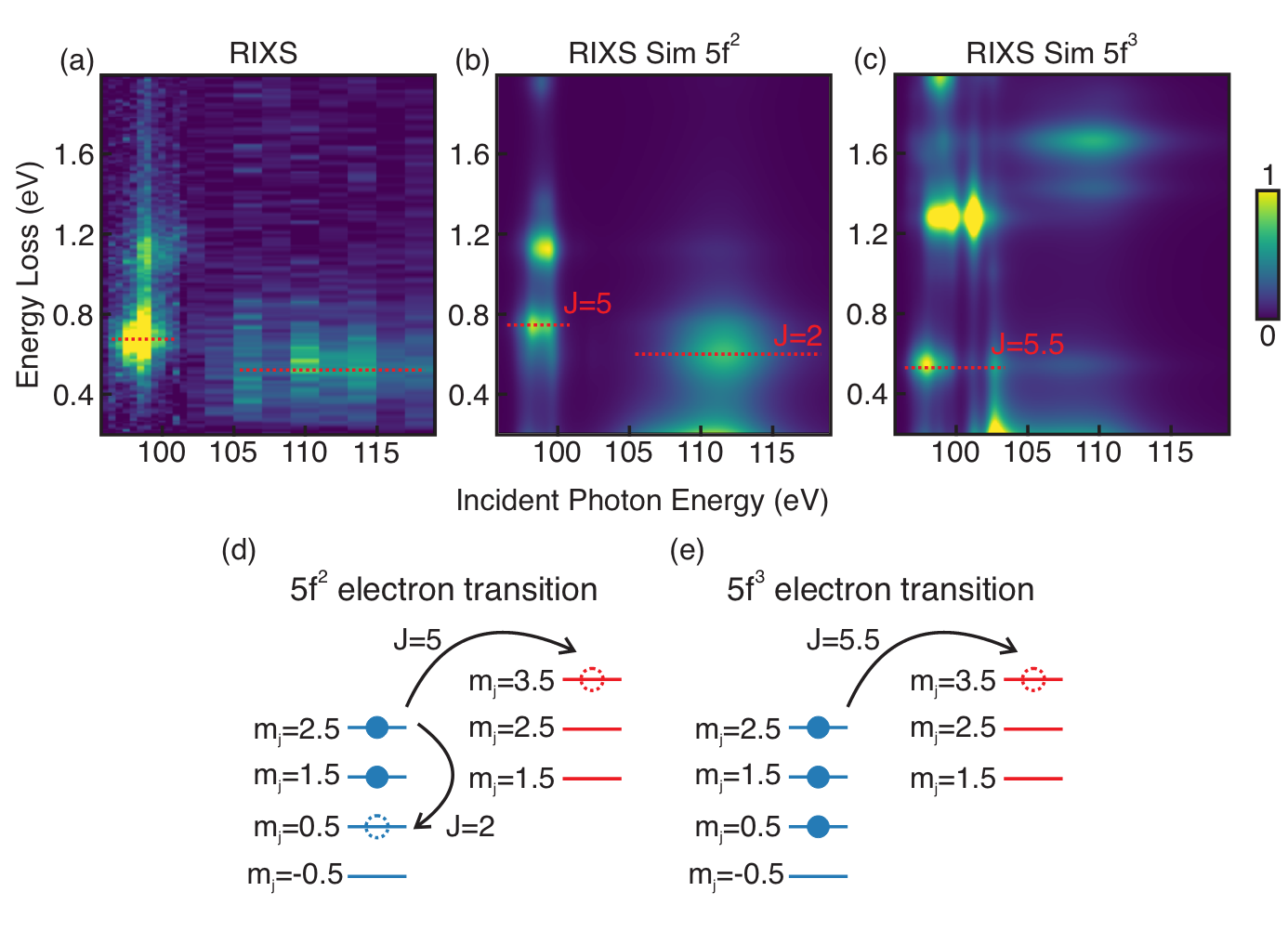}
    \caption{Atomic multiplet excitation spectrum. (a-c) RIXS spectra for UTe$_2$ and multiplet simulations for 5$f^2$ and 5$f^3$ configurations. Total angular momentum (J) is indicated in red for key excitation features of the simulations. (d) Electronic transitions associated with excitation from the 5$f^2$ ground state (J=4) to excited states with J=2 and 5 angular momentum. Single-electron states with j=5/2 are labeled in blue, and with j=7/2 are labeled in red. (e) A diagram shows the primary electronic transition associated with the low energy J=5.5 excitation of 5$f^3$ atoms.}
    \label{fig:fig2}
\end{figure*}

% 4/5th paragraph Fig. 2...
A deeper understanding of the electronic structure can be obtained by using RIXS to observe final states that are projected into following the photon-mediated decay of a core hole resonance state \cite{rixs_review}. The RIXS spectrum after removing the elastic peak is presented in Fig. \ref{fig:fig2}(a) together with multiplet simulations for 5$f^2$ and 5$f^3$ scenarios (panels b-c). The total angular momentum quantum number J is used to label low energy excitations, as the the energy scale of intra-atomic j-j coupling is larger than that of crystal field splitting.

A one-to-one correspondence of features can be easily identified between the data and 5$f^2$ simulation (Fig. \ref{fig:fig2}(a-b)), and the measurement closely resembles the 5$f^2$-derived spectrum of URu$_2$Si$_2$ \cite{wray2015}. Both the simulation and data feature two excitations spaced apart by $\sim$0.2 eV at the h$\nu\sim$99 eV resonance and one energetically distinct excitation at the h$\nu\sim110$ eV resonance. The key factor distinguishing the 5$f^3$ scenario is that regardless of the choice of modeling parameters, there is only one low energy spectral feature with energy E$<$1 eV (labeled J=5.5). The principal f-electron transitions are shown in \ref{fig:fig2}(d-e) for 5$f^2$ and 5$f^3$ scenarios. In the 5$f^2$ configuration, the lowest two excitations have angular momentum J=2 ($\sim$0.6 eV) and J=5 ($\sim$0.73 eV), and can be created through $j=\frac{5}{2}\xrightarrow{}\frac{5}{2}$ and $j=\frac{5}{2}\xrightarrow{}\frac{7}{2}$ single electron transitions. The higher energy E=1.1 eV excitation has J=4 ($^3G_4$ multiplet symmetry), but is not a focus of this investigation. For 5$f^3$, the $<1$eV sector contains just one excitation excited through a $j=\frac{5}{2}\xrightarrow{}\frac{7}{2}$ transition, as transitions within the $j=\frac{5}{2}$ manifold create anti-parallel electron spins, which pushes excitation energy to $>$1 eV. (see supplement for further symmetry details \cite{SI}) The large $\sim$130 meV energy difference between these features at the low- and high-energy resonances is therefore strongly indicative of a 5$f^2$-based electronic structure. In a itinerant picture the J=2 excitation is a single-particle transition within the $j=\frac{5}{2}$ state manifold, and thus is expected to be nearly gapless. The observed $\sim$0.6 eV excitation energy matches expectations from Hund's rule coupling, and is confirmation of the locally correlated nature of the f-electrons occupying uranium.

\begin{figure}
    \centering
    \includegraphics[width=0.5\textwidth]{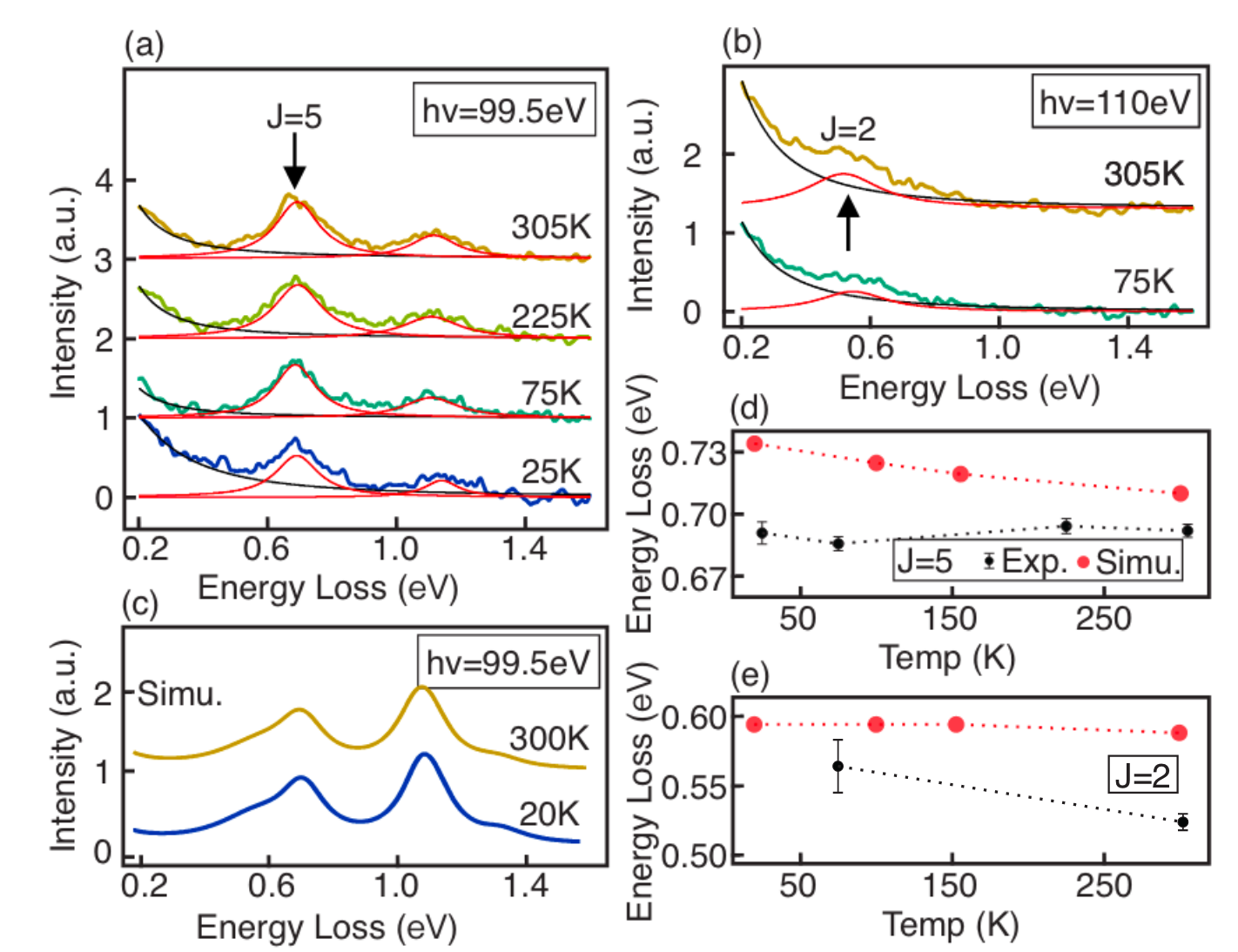}
    \caption{Magnetic coupling in a uranium dimer. (a) Energy of the J=5 excitation is fitted as a a Lorentzian function for temperature-resolved RIXS measurements at h$\nu$=99.5 eV. A single Lorentzian peak fixed at zero energy loss is used to account for extended tail of quasielastic scattering. (b) The J=2 excitation is fitted from RIXS scans at h$\nu$=110 eV. (c) Multiplet simulations of the 5$f^2$ h$\nu$=99.5 eV RIXS spectrum. (d-e, black) The temperature dependence of J=5 and J=2 excitation energies is shown with standard deviation fitting error, and compared with (red) peak energies from the multiplet simulation.}
    \label{fig:fig3}
\end{figure}

Energy level splitting from the magnetic exchange interactions between dimerized uranium atoms is significant for understanding the Cooper pairing mechanism \cite{yang_paper,qimiao_paper,anderson_model}, but is too small to resolve directly in the RIXS spectra. However, other consequences of magnetism are resolvable within simulations. Strong magnetic alignment within a dimer will increase the energy gap between the ground state and the J=5 excitation, as the associated $j=\frac{5}{2}\xrightarrow{}\frac{7}{2}$ transition flips the spin of one electron (see Fig. \ref{fig:fig2}(d)). The J=5.5 5$f^3$ excitation also occurs via this transition and has nearly identical temperature dependence to J=5, but is not separately considered in Fig. \ref{fig:fig3}. Magnetic alignment of the simulated dimer atoms is maximized at low temperature, resulting in a 25 meV greater energy cost to create this excitation at T=20K compared with room temperature (see Fig. \ref{fig:fig3}(d)).

%together with corresponding 5$f^2$ simulations in \ref{fig:fig3}(c). The energies and intensities of the prominent J=5 peak and less-visible J=2 peak are similar to previous reported uranium compound RIXS results \cite{wray2015}. 

%The fittings are conducted for all relevant temperatures, however they are not sufficiently accurate to observe the predicted temperature trend of the J=2 feature, due to the poor signal-to-background ratio (see supplement).  

% The downward trend of excitation energy as temperature increases in the J=5 simulation has a natural explanation based on a localized picture. The j=$\frac{7}{2}$ electron in the J=5 excited state has a greatly reduced spin moment relative to the J=4 ground state. Because magnetic exchange interactions occur in the spin sector, this reduction in total spin moment comes at a cost with respect to the magnetic interaction energy of the uranium dimer. At higher temperatures thermal fluctuations cause the local moment vectors of dimer-paired uranium atoms to be increasingly misaligned, making it cost less energy to reduce the spin moment on one of them.

Lorentzian fits of the experimentally observed J=5 and J=2 features are shown in Fig. \ref{fig:fig3}(a-b), and the temperature dependence of feature energies is summarized in Fig. \ref{fig:fig3}(d-e). The observed energy difference across the temperature range is $E(20K)-E(300K)=-1 \pm 3$ meV, which is incompatible with the predicted value of 25 meV. The trend for J=2 cannot be quantitatively compared due to contamination from the elastic line, which contributes to large error bars. As with the weak experimental XAS linear dichroism signal, a likely reason for the lack of temperature dependence in the J=5 feature energy is that scattering with itinerant 6$d$ electrons causes the low temperature symmetries to be less cleanly defined than the multiplet model predicts, with weaker alignment of angular momentum between the dimerized atoms. Differences in the atomic coordinates between high and low temperature may also be a factor. Regardless of the reason, the result suggests that intra-dimer magnetic interactions play a much less determinative role in local energetics and symmetries than is expected from a from a minimal local picture that combines atomic physics and the large predicted $\sim$50 meV intra-dimer exchange interaction \cite{yang_paper}.

Given the close correspondence of these resonant spectroscopy data with a 5$f^2$-based picture, it is important to review how this can be understood with respect to the quasiparticles observed by angle-resolved photoemission spectroscopy (ARPES), which have also been interpreted as providing definitive evidence for a 5$f^2$6$d^1$-based picture \cite{miao_paper}. Specifically, ARPES measurements have observed a highly itinerant band associated with the Te dimers, which results in a Luttinger-based electron count of $n_{Te}\sim$11.0 for the Te atoms (Te$^{-1.5}$ valence). Furthermore, a dispersive uranium 6$d$ band with a Luttinger count of $\sim$0.8 has also been observed (see supplementary material \cite{SI}), and has nonzero dispersion along the surface-normal axis indicating bulk character \cite{miao_paper}. These feature attributions are consistently interpreted with core level resonance, de Haas van Alphen measurements \cite{UTe2_dHvA}, DFT+U \cite{anderson_model}, and DMFT \cite{miao_paper, Armitage_optical}, and combine to require a roughly 5$f^{2.2\pm \delta}$6$d^{0.8\mp \delta}$ picture, where one expects $|\delta|\lesssim 0.1$. Deviating from this picture requires a significant re-interpretation of the U 6$d$ and Te 5$p$ band dispersions, such that the occupied k-space volume beneath the Fermi level is much smaller than observed. These attributions of low energy quasiparticles and local degrees of freedom are firmly corroborated by the 5$f^2$-like multiplet structure observed with RIXS and XAS, and provide foundational ingredients for the construction of low energy models.

Proposals of a 5$f^3$-like valence state stem primarily from numerical and experimental estimates of uranium \emph{charge density} \cite{fujimori_paper,fujimori_paper4,UTe2_dHvA}, with the simulations in Ref. \cite{fujimori_paper} yielding an $f$-electron density that approaches 3 ($n_f=2.73$). However, it is important to remember that the orbital-resolved charge density and effective valence state are not at all equivalent, and are rarely in close agreement for strongly correlated materials. This is due to the role of local hybridization in expanding the effective orbitals of a cluster model, which is termed the nephelauxetic effect when discussing transition metals. For example, the Mott insulator NiO is a model 3$d^8$ material, but has a 3d electron count of $n_d \sim8.2$ \cite{Elp_NiO} The difference tends to be significantly greater in cases like that of UTe$_2$, where the strongly correlated element has a mostly unfilled valence orbital manifold and ligands are electron rich. An example of this limit is the 3d$^1$ correlated insulator VO$_2$ which has a 3$d$ electron number that rounds to $n_d\sim2$ \cite{VO2_paper}. A similar scenario with 5$f^3$-like charge density and 5$f^2$-like multiplet symmetry is found for the electron rich compound UFe$_2$Si$_2$ \cite{amorese_paper}. In this context, the recent DFT+U(ED) model prediction \cite{fujimori_paper} of $n_f=2.73$ f-electron number for UTe$_2$ can only be taken to imply an effective valence configuration of 5$f^{2+\delta}$6$d^{1- \delta}$ with $\delta<0.7$ serving as an extreme upper limit, not inconsistent with the ARPES-based picture of 5$f^{2.2}$6$d^{0.8}$ effective valence.

In conclusion, we have shown that multiplet-resolving XAS and RIXS spectra at the uranium O-edge are strongly consistent with expectations for a 5$f^2$6$d^1$ effective valence configuration, and inconsistent with the alternative 5$f^3$ scenario. The RIXS spectrum reveals a gapped J=2 excitation that implies strong Hund's rule alignment on uranium, and has no analogue in the 5$f^3$ multiplet excitation spectrum. The amplitude and temperature dependence of XAS linear dichroism as well as the temperature dependence in RIXS excitation energies are all found to be weaker than expected for the magnetically interacting uranium dimer. These observations suggest that additional factors must be considered to understand the uranium dimer symmetries and intra-dimer spin correlations that may be significant in defining the local environment for triplet Cooper pairing. In particular, scattering from atomic 5$f$-6$d$ electron interactions is expected to reduce the amplitude of both dichroism and temperature dependence, and is proposed to be a significant factor.

\emph{Acknowledgements:} This research used resources of the Advanced Light Source, a U.S. DOE Office of Science User Facility under Contract No. DE-AC02-05CH11231. Research at New York University was supported by the National Science Foundation under Grant No. DMR-2105081. Research at the University of Maryland was supported by Department of Energy Award No. DE-SC-0019154 (transport experiments) and the Gordon and Betty Moore Foundation EPiQS Initiative through Grant No. GBMF9071 (materials synthesis), NIST, and the Maryland Quantum Materials Center.

\bibliography{ref}

\begin{thebibliography}{34}%
\makeatletter
\providecommand \@ifxundefined [1]{%
 \@ifx{#1\undefined}
}%
\providecommand \@ifnum [1]{%
 \ifnum #1\expandafter \@firstoftwo
 \else \expandafter \@secondoftwo
 \fi
}%
\providecommand \@ifx [1]{%
 \ifx #1\expandafter \@firstoftwo
 \else \expandafter \@secondoftwo
 \fi
}%
\providecommand \natexlab [1]{#1}%
\providecommand \enquote  [1]{``#1''}%
\providecommand \bibnamefont  [1]{#1}%
\providecommand \bibfnamefont [1]{#1}%
\providecommand \citenamefont [1]{#1}%
\providecommand \href@noop [0]{\@secondoftwo}%
\providecommand \href [0]{\begingroup \@sanitize@url \@href}%
\providecommand \@href[1]{\@@startlink{#1}\@@href}%
\providecommand \@@href[1]{\endgroup#1\@@endlink}%
\providecommand \@sanitize@url [0]{\catcode `\\12\catcode `\$12\catcode
  `\&12\catcode `\#12\catcode `\^12\catcode `\_12\catcode `\%12\relax}%
\providecommand \@@startlink[1]{}%
\providecommand \@@endlink[0]{}%
\providecommand \url  [0]{\begingroup\@sanitize@url \@url }%
\providecommand \@url [1]{\endgroup\@href {#1}{\urlprefix }}%
\providecommand \urlprefix  [0]{URL }%
\providecommand \Eprint [0]{\href }%
\providecommand \doibase [0]{http://dx.doi.org/}%
\providecommand \selectlanguage [0]{\@gobble}%
\providecommand \bibinfo  [0]{\@secondoftwo}%
\providecommand \bibfield  [0]{\@secondoftwo}%
\providecommand \translation [1]{[#1]}%
\providecommand \BibitemOpen [0]{}%
\providecommand \bibitemStop [0]{}%
\providecommand \bibitemNoStop [0]{.\EOS\space}%
\providecommand \EOS [0]{\spacefactor3000\relax}%
\providecommand \BibitemShut  [1]{\csname bibitem#1\endcsname}%
\let\auto@bib@innerbib\@empty
%</preamble>
\bibitem [{\citenamefont {Ran}\ \emph {et~al.}(2019{\natexlab{a}})\citenamefont
  {Ran}, \citenamefont {Eckberg}, \citenamefont {Ding}, \citenamefont
  {Furukawa}, \citenamefont {Metz}, \citenamefont {Saha}, \citenamefont {Liu},
  \citenamefont {Zic}, \citenamefont {Kim}, \citenamefont {Paglione} \emph
  {et~al.}}]{ranshen_early_paper}%
  \BibitemOpen
  \bibfield  {author} {\bibinfo {author} {\bibfnamefont {S.}~\bibnamefont
  {Ran}}, \bibinfo {author} {\bibfnamefont {C.}~\bibnamefont {Eckberg}},
  \bibinfo {author} {\bibfnamefont {Q.-P.}\ \bibnamefont {Ding}}, \bibinfo
  {author} {\bibfnamefont {Y.}~\bibnamefont {Furukawa}}, \bibinfo {author}
  {\bibfnamefont {T.}~\bibnamefont {Metz}}, \bibinfo {author} {\bibfnamefont
  {S.~R.}\ \bibnamefont {Saha}}, \bibinfo {author} {\bibfnamefont {I.-L.}\
  \bibnamefont {Liu}}, \bibinfo {author} {\bibfnamefont {M.}~\bibnamefont
  {Zic}}, \bibinfo {author} {\bibfnamefont {H.}~\bibnamefont {Kim}}, \bibinfo
  {author} {\bibfnamefont {J.}~\bibnamefont {Paglione}},  \emph {et~al.},\
  }\href@noop {} {\bibfield  {journal} {\bibinfo  {journal} {Science}\ }\textbf
  {\bibinfo {volume} {365}},\ \bibinfo {pages} {684} (\bibinfo {year}
  {2019}{\natexlab{a}})}\BibitemShut {NoStop}%
\bibitem [{\citenamefont {Ran}\ \emph {et~al.}(2019{\natexlab{b}})\citenamefont
  {Ran}, \citenamefont {Liu}, \citenamefont {Eo}, \citenamefont {Campbell},
  \citenamefont {Neves}, \citenamefont {Fuhrman}, \citenamefont {Saha},
  \citenamefont {Eckberg}, \citenamefont {Kim}, \citenamefont {Graf} \emph
  {et~al.}}]{Ute2_upper_critical_field}%
  \BibitemOpen
  \bibfield  {author} {\bibinfo {author} {\bibfnamefont {S.}~\bibnamefont
  {Ran}}, \bibinfo {author} {\bibfnamefont {I.-L.}\ \bibnamefont {Liu}},
  \bibinfo {author} {\bibfnamefont {Y.~S.}\ \bibnamefont {Eo}}, \bibinfo
  {author} {\bibfnamefont {D.~J.}\ \bibnamefont {Campbell}}, \bibinfo {author}
  {\bibfnamefont {P.~M.}\ \bibnamefont {Neves}}, \bibinfo {author}
  {\bibfnamefont {W.~T.}\ \bibnamefont {Fuhrman}}, \bibinfo {author}
  {\bibfnamefont {S.~R.}\ \bibnamefont {Saha}}, \bibinfo {author}
  {\bibfnamefont {C.}~\bibnamefont {Eckberg}}, \bibinfo {author} {\bibfnamefont
  {H.}~\bibnamefont {Kim}}, \bibinfo {author} {\bibfnamefont {D.}~\bibnamefont
  {Graf}},  \emph {et~al.},\ }\href@noop {} {\bibfield  {journal} {\bibinfo
  {journal} {Nature physics}\ }\textbf {\bibinfo {volume} {15}},\ \bibinfo
  {pages} {1250} (\bibinfo {year} {2019}{\natexlab{b}})}\BibitemShut {NoStop}%
\bibitem [{\citenamefont {Wei}\ \emph {et~al.}(2022)\citenamefont {Wei},
  \citenamefont {Saykin}, \citenamefont {Miller}, \citenamefont {Ran},
  \citenamefont {Saha}, \citenamefont {Agterberg}, \citenamefont {Schmalian},
  \citenamefont {Butch}, \citenamefont {Paglione},\ and\ \citenamefont
  {Kapitulnik}}]{Ute2_TRSB}%
  \BibitemOpen
  \bibfield  {author} {\bibinfo {author} {\bibfnamefont {D.~S.}\ \bibnamefont
  {Wei}}, \bibinfo {author} {\bibfnamefont {D.}~\bibnamefont {Saykin}},
  \bibinfo {author} {\bibfnamefont {O.~Y.}\ \bibnamefont {Miller}}, \bibinfo
  {author} {\bibfnamefont {S.}~\bibnamefont {Ran}}, \bibinfo {author}
  {\bibfnamefont {S.~R.}\ \bibnamefont {Saha}}, \bibinfo {author}
  {\bibfnamefont {D.~F.}\ \bibnamefont {Agterberg}}, \bibinfo {author}
  {\bibfnamefont {J.}~\bibnamefont {Schmalian}}, \bibinfo {author}
  {\bibfnamefont {N.~P.}\ \bibnamefont {Butch}}, \bibinfo {author}
  {\bibfnamefont {J.}~\bibnamefont {Paglione}}, \ and\ \bibinfo {author}
  {\bibfnamefont {A.}~\bibnamefont {Kapitulnik}},\ }\href {\doibase
  10.1103/physrevb.105.024521} {\bibfield  {journal} {\bibinfo  {journal}
  {Physical Review B}\ }\textbf {\bibinfo {volume} {105}},\ \bibinfo {pages}
  {024521} (\bibinfo {year} {2022})}\BibitemShut {NoStop}%
\bibitem [{\citenamefont {Nayak}\ \emph {et~al.}(2008)\citenamefont {Nayak},
  \citenamefont {Simon}, \citenamefont {Stern}, \citenamefont {Freedman},\ and\
  \citenamefont {Das~Sarma}}]{majorana}%
  \BibitemOpen
  \bibfield  {author} {\bibinfo {author} {\bibfnamefont {C.}~\bibnamefont
  {Nayak}}, \bibinfo {author} {\bibfnamefont {S.~H.}\ \bibnamefont {Simon}},
  \bibinfo {author} {\bibfnamefont {A.}~\bibnamefont {Stern}}, \bibinfo
  {author} {\bibfnamefont {M.}~\bibnamefont {Freedman}}, \ and\ \bibinfo
  {author} {\bibfnamefont {S.}~\bibnamefont {Das~Sarma}},\ }\href {\doibase
  10.1103/RevModPhys.80.1083} {\bibfield  {journal} {\bibinfo  {journal} {Rev.
  Mod. Phys.}\ }\textbf {\bibinfo {volume} {80}},\ \bibinfo {pages} {1083}
  (\bibinfo {year} {2008})}\BibitemShut {NoStop}%
\bibitem [{\citenamefont {Xu}\ \emph {et~al.}(2019)\citenamefont {Xu},
  \citenamefont {Sheng},\ and\ \citenamefont {Yang}}]{yang_paper}%
  \BibitemOpen
  \bibfield  {author} {\bibinfo {author} {\bibfnamefont {Y.}~\bibnamefont
  {Xu}}, \bibinfo {author} {\bibfnamefont {Y.}~\bibnamefont {Sheng}}, \ and\
  \bibinfo {author} {\bibfnamefont {Y.-f.}\ \bibnamefont {Yang}},\ }\href
  {\doibase 10.1103/PhysRevLett.123.217002} {\bibfield  {journal} {\bibinfo
  {journal} {Phys. Rev. Lett.}\ }\textbf {\bibinfo {volume} {123}},\ \bibinfo
  {pages} {217002} (\bibinfo {year} {2019})}\BibitemShut {NoStop}%
\bibitem [{\citenamefont {Chen}\ \emph {et~al.}(2021)\citenamefont {Chen},
  \citenamefont {Hu}, \citenamefont {Lane}, \citenamefont {Nica}, \citenamefont
  {Zhu},\ and\ \citenamefont {Si}}]{qimiao_paper}%
  \BibitemOpen
  \bibfield  {author} {\bibinfo {author} {\bibfnamefont {L.}~\bibnamefont
  {Chen}}, \bibinfo {author} {\bibfnamefont {H.}~\bibnamefont {Hu}}, \bibinfo
  {author} {\bibfnamefont {C.}~\bibnamefont {Lane}}, \bibinfo {author}
  {\bibfnamefont {E.~M.}\ \bibnamefont {Nica}}, \bibinfo {author}
  {\bibfnamefont {J.-X.}\ \bibnamefont {Zhu}}, \ and\ \bibinfo {author}
  {\bibfnamefont {Q.}~\bibnamefont {Si}},\ }\href {\doibase
  10.48550/ARXIV.2112.14750} {\enquote {\bibinfo {title} {Multiorbital
  spin-triplet pairing and spin resonance in the heavy-fermion superconductor
  $\mathrm{UTe_2}$},}\ } (\bibinfo {year} {2021})\BibitemShut {NoStop}%
\bibitem [{\citenamefont {Ishizuka}\ and\ \citenamefont
  {Yanase}(2021)}]{anderson_model}%
  \BibitemOpen
  \bibfield  {author} {\bibinfo {author} {\bibfnamefont {J.}~\bibnamefont
  {Ishizuka}}\ and\ \bibinfo {author} {\bibfnamefont {Y.}~\bibnamefont
  {Yanase}},\ }\href {\doibase 10.1103/PhysRevB.103.094504} {\bibfield
  {journal} {\bibinfo  {journal} {Phys. Rev. B}\ }\textbf {\bibinfo {volume}
  {103}},\ \bibinfo {pages} {094504} (\bibinfo {year} {2021})}\BibitemShut
  {NoStop}%
\bibitem [{\citenamefont {Fujimori}\ \emph {et~al.}(2021)\citenamefont
  {Fujimori}, \citenamefont {Kawasaki}, \citenamefont {Takeda}, \citenamefont
  {Yamagami}, \citenamefont {Nakamura}, \citenamefont {Homma},\ and\
  \citenamefont {Aoki}}]{fujimori_paper4}%
  \BibitemOpen
  \bibfield  {author} {\bibinfo {author} {\bibfnamefont {S.-i.}\ \bibnamefont
  {Fujimori}}, \bibinfo {author} {\bibfnamefont {I.}~\bibnamefont {Kawasaki}},
  \bibinfo {author} {\bibfnamefont {Y.}~\bibnamefont {Takeda}}, \bibinfo
  {author} {\bibfnamefont {H.}~\bibnamefont {Yamagami}}, \bibinfo {author}
  {\bibfnamefont {A.}~\bibnamefont {Nakamura}}, \bibinfo {author}
  {\bibfnamefont {Y.}~\bibnamefont {Homma}}, \ and\ \bibinfo {author}
  {\bibfnamefont {D.}~\bibnamefont {Aoki}},\ }\href@noop {} {\bibfield
  {journal} {\bibinfo  {journal} {Journal of the Physical Society of Japan}\
  }\textbf {\bibinfo {volume} {90}},\ \bibinfo {pages} {015002} (\bibinfo
  {year} {2021})}\BibitemShut {NoStop}%
\bibitem [{\citenamefont {Miao}\ \emph {et~al.}(2020)\citenamefont {Miao},
  \citenamefont {Liu}, \citenamefont {Xu}, \citenamefont {Kotta}, \citenamefont
  {Kang}, \citenamefont {Ran}, \citenamefont {Paglione}, \citenamefont
  {Kotliar}, \citenamefont {Butch}, \citenamefont {Denlinger} \emph
  {et~al.}}]{miao_paper}%
  \BibitemOpen
  \bibfield  {author} {\bibinfo {author} {\bibfnamefont {L.}~\bibnamefont
  {Miao}}, \bibinfo {author} {\bibfnamefont {S.}~\bibnamefont {Liu}}, \bibinfo
  {author} {\bibfnamefont {Y.}~\bibnamefont {Xu}}, \bibinfo {author}
  {\bibfnamefont {E.~C.}\ \bibnamefont {Kotta}}, \bibinfo {author}
  {\bibfnamefont {C.-J.}\ \bibnamefont {Kang}}, \bibinfo {author}
  {\bibfnamefont {S.}~\bibnamefont {Ran}}, \bibinfo {author} {\bibfnamefont
  {J.}~\bibnamefont {Paglione}}, \bibinfo {author} {\bibfnamefont
  {G.}~\bibnamefont {Kotliar}}, \bibinfo {author} {\bibfnamefont {N.~P.}\
  \bibnamefont {Butch}}, \bibinfo {author} {\bibfnamefont {J.~D.}\ \bibnamefont
  {Denlinger}},  \emph {et~al.},\ }\href@noop {} {\bibfield  {journal}
  {\bibinfo  {journal} {Physical review letters}\ }\textbf {\bibinfo {volume}
  {124}},\ \bibinfo {pages} {076401} (\bibinfo {year} {2020})}\BibitemShut
  {NoStop}%
\bibitem [{\citenamefont {Thomas}\ \emph {et~al.}(2020)\citenamefont {Thomas},
  \citenamefont {Santos}, \citenamefont {Christensen}, \citenamefont {Asaba},
  \citenamefont {Ronning}, \citenamefont {Thompson}, \citenamefont {Bauer},
  \citenamefont {Fernandes}, \citenamefont {Fabbris},\ and\ \citenamefont
  {Rosa}}]{charge_density_resolved_paper}%
  \BibitemOpen
  \bibfield  {author} {\bibinfo {author} {\bibfnamefont {S.~M.}\ \bibnamefont
  {Thomas}}, \bibinfo {author} {\bibfnamefont {F.~B.}\ \bibnamefont {Santos}},
  \bibinfo {author} {\bibfnamefont {M.~H.}\ \bibnamefont {Christensen}},
  \bibinfo {author} {\bibfnamefont {T.}~\bibnamefont {Asaba}}, \bibinfo
  {author} {\bibfnamefont {F.}~\bibnamefont {Ronning}}, \bibinfo {author}
  {\bibfnamefont {J.~D.}\ \bibnamefont {Thompson}}, \bibinfo {author}
  {\bibfnamefont {E.~D.}\ \bibnamefont {Bauer}}, \bibinfo {author}
  {\bibfnamefont {R.~M.}\ \bibnamefont {Fernandes}}, \bibinfo {author}
  {\bibfnamefont {G.}~\bibnamefont {Fabbris}}, \ and\ \bibinfo {author}
  {\bibfnamefont {P.~F.~S.}\ \bibnamefont {Rosa}},\ }\href {\doibase
  10.1126/sciadv.abc8709} {\bibfield  {journal} {\bibinfo  {journal} {Science
  Advances}\ }\textbf {\bibinfo {volume} {6}},\ \bibinfo {pages} {eabc8709}
  (\bibinfo {year} {2020})}\BibitemShut {NoStop}%
\bibitem [{\citenamefont {Aoki}\ \emph {et~al.}(2022)\citenamefont {Aoki},
  \citenamefont {Sakai}, \citenamefont {Opletal}, \citenamefont {Tokiwa},
  \citenamefont {Ishizuka}, \citenamefont {Yanase}, \citenamefont {Harima},
  \citenamefont {Nakamura}, \citenamefont {Li}, \citenamefont {Homma},
  \citenamefont {Shimazu}, \citenamefont {Knebel}, \citenamefont {Flouquet},\
  and\ \citenamefont {Haga}}]{UTe2_dHvA}%
  \BibitemOpen
  \bibfield  {author} {\bibinfo {author} {\bibfnamefont {D.}~\bibnamefont
  {Aoki}}, \bibinfo {author} {\bibfnamefont {H.}~\bibnamefont {Sakai}},
  \bibinfo {author} {\bibfnamefont {P.}~\bibnamefont {Opletal}}, \bibinfo
  {author} {\bibfnamefont {Y.}~\bibnamefont {Tokiwa}}, \bibinfo {author}
  {\bibfnamefont {J.}~\bibnamefont {Ishizuka}}, \bibinfo {author}
  {\bibfnamefont {Y.}~\bibnamefont {Yanase}}, \bibinfo {author} {\bibfnamefont
  {H.}~\bibnamefont {Harima}}, \bibinfo {author} {\bibfnamefont
  {A.}~\bibnamefont {Nakamura}}, \bibinfo {author} {\bibfnamefont
  {D.}~\bibnamefont {Li}}, \bibinfo {author} {\bibfnamefont {Y.}~\bibnamefont
  {Homma}}, \bibinfo {author} {\bibfnamefont {Y.}~\bibnamefont {Shimazu}},
  \bibinfo {author} {\bibfnamefont {G.}~\bibnamefont {Knebel}}, \bibinfo
  {author} {\bibfnamefont {J.}~\bibnamefont {Flouquet}}, \ and\ \bibinfo
  {author} {\bibfnamefont {Y.}~\bibnamefont {Haga}},\ }\href {\doibase
  https://doi.org/10.7566/JPSJ.91.083704} {\bibfield  {journal} {\bibinfo
  {journal} {J. Phys. Soc. J.}\ }\textbf {\bibinfo {volume} {91}},\ \bibinfo
  {pages} {083704} (\bibinfo {year} {2022})}\BibitemShut {NoStop}%
\bibitem [{\citenamefont {Ikeda}\ \emph {et~al.}(2006)\citenamefont {Ikeda},
  \citenamefont {Sakai}, \citenamefont {Aoki}, \citenamefont {Homma},
  \citenamefont {Yamamoto}, \citenamefont {Nakamura}, \citenamefont {Shiokawa},
  \citenamefont {Haga},\ and\ \citenamefont {{\=O}nuki}}]{Ute2_structure}%
  \BibitemOpen
  \bibfield  {author} {\bibinfo {author} {\bibfnamefont {S.}~\bibnamefont
  {Ikeda}}, \bibinfo {author} {\bibfnamefont {H.}~\bibnamefont {Sakai}},
  \bibinfo {author} {\bibfnamefont {D.}~\bibnamefont {Aoki}}, \bibinfo {author}
  {\bibfnamefont {Y.}~\bibnamefont {Homma}}, \bibinfo {author} {\bibfnamefont
  {E.}~\bibnamefont {Yamamoto}}, \bibinfo {author} {\bibfnamefont
  {A.}~\bibnamefont {Nakamura}}, \bibinfo {author} {\bibfnamefont
  {Y.}~\bibnamefont {Shiokawa}}, \bibinfo {author} {\bibfnamefont
  {Y.}~\bibnamefont {Haga}}, \ and\ \bibinfo {author} {\bibfnamefont
  {Y.}~\bibnamefont {{\=O}nuki}},\ }\href@noop {} {\bibfield  {journal}
  {\bibinfo  {journal} {journal of the physical society of japan}\ }\textbf
  {\bibinfo {volume} {75}},\ \bibinfo {pages} {116} (\bibinfo {year}
  {2006})}\BibitemShut {NoStop}%
\bibitem [{\citenamefont {Saxena}\ \emph {et~al.}(2000)\citenamefont {Saxena},
  \citenamefont {Agarwal}, \citenamefont {Ahilan}, \citenamefont {Grosche},
  \citenamefont {Haselwimmer}, \citenamefont {Steiner}, \citenamefont {Pugh},
  \citenamefont {Walker}, \citenamefont {Julian}, \citenamefont {Monthoux}
  \emph {et~al.}}]{UGe2_paper}%
  \BibitemOpen
  \bibfield  {author} {\bibinfo {author} {\bibfnamefont {S.}~\bibnamefont
  {Saxena}}, \bibinfo {author} {\bibfnamefont {P.}~\bibnamefont {Agarwal}},
  \bibinfo {author} {\bibfnamefont {K.}~\bibnamefont {Ahilan}}, \bibinfo
  {author} {\bibfnamefont {F.}~\bibnamefont {Grosche}}, \bibinfo {author}
  {\bibfnamefont {R.}~\bibnamefont {Haselwimmer}}, \bibinfo {author}
  {\bibfnamefont {M.}~\bibnamefont {Steiner}}, \bibinfo {author} {\bibfnamefont
  {E.}~\bibnamefont {Pugh}}, \bibinfo {author} {\bibfnamefont {I.}~\bibnamefont
  {Walker}}, \bibinfo {author} {\bibfnamefont {S.}~\bibnamefont {Julian}},
  \bibinfo {author} {\bibfnamefont {P.}~\bibnamefont {Monthoux}},  \emph
  {et~al.},\ }\href@noop {} {\bibfield  {journal} {\bibinfo  {journal}
  {Nature}\ }\textbf {\bibinfo {volume} {406}},\ \bibinfo {pages} {587}
  (\bibinfo {year} {2000})}\BibitemShut {NoStop}%
\bibitem [{\citenamefont {Aoki}\ \emph {et~al.}(2001)\citenamefont {Aoki},
  \citenamefont {Huxley}, \citenamefont {Ressouche}, \citenamefont
  {Braithwaite}, \citenamefont {Flouquet}, \citenamefont {Brison},
  \citenamefont {Lhotel},\ and\ \citenamefont {Paulsen}}]{URhGe_paper}%
  \BibitemOpen
  \bibfield  {author} {\bibinfo {author} {\bibfnamefont {D.}~\bibnamefont
  {Aoki}}, \bibinfo {author} {\bibfnamefont {A.}~\bibnamefont {Huxley}},
  \bibinfo {author} {\bibfnamefont {E.}~\bibnamefont {Ressouche}}, \bibinfo
  {author} {\bibfnamefont {D.}~\bibnamefont {Braithwaite}}, \bibinfo {author}
  {\bibfnamefont {J.}~\bibnamefont {Flouquet}}, \bibinfo {author}
  {\bibfnamefont {J.-P.}\ \bibnamefont {Brison}}, \bibinfo {author}
  {\bibfnamefont {E.}~\bibnamefont {Lhotel}}, \ and\ \bibinfo {author}
  {\bibfnamefont {C.}~\bibnamefont {Paulsen}},\ }\href@noop {} {\bibfield
  {journal} {\bibinfo  {journal} {Nature}\ }\textbf {\bibinfo {volume} {413}},\
  \bibinfo {pages} {613} (\bibinfo {year} {2001})}\BibitemShut {NoStop}%
\bibitem [{\citenamefont {Huy}\ \emph {et~al.}(2007)\citenamefont {Huy},
  \citenamefont {Gasparini}, \citenamefont {de~Nijs}, \citenamefont {Huang},
  \citenamefont {Klaasse}, \citenamefont {Gortenmulder}, \citenamefont
  {de~Visser}, \citenamefont {Hamann}, \citenamefont {G{\"o}rlach},\ and\
  \citenamefont {L{\"o}hneysen}}]{UCoGe_paper}%
  \BibitemOpen
  \bibfield  {author} {\bibinfo {author} {\bibfnamefont {N.~T.}\ \bibnamefont
  {Huy}}, \bibinfo {author} {\bibfnamefont {A.}~\bibnamefont {Gasparini}},
  \bibinfo {author} {\bibfnamefont {D.~E.}\ \bibnamefont {de~Nijs}}, \bibinfo
  {author} {\bibfnamefont {Y.}~\bibnamefont {Huang}}, \bibinfo {author}
  {\bibfnamefont {J.~C.~P.}\ \bibnamefont {Klaasse}}, \bibinfo {author}
  {\bibfnamefont {T.}~\bibnamefont {Gortenmulder}}, \bibinfo {author}
  {\bibfnamefont {A.}~\bibnamefont {de~Visser}}, \bibinfo {author}
  {\bibfnamefont {A.}~\bibnamefont {Hamann}}, \bibinfo {author} {\bibfnamefont
  {T.}~\bibnamefont {G{\"o}rlach}}, \ and\ \bibinfo {author} {\bibfnamefont
  {H.~V.}\ \bibnamefont {L{\"o}hneysen}},\ }\href@noop {} {\bibfield  {journal}
  {\bibinfo  {journal} {Physical review letters}\ }\textbf {\bibinfo {volume}
  {99}},\ \bibinfo {pages} {067006} (\bibinfo {year} {2007})}\BibitemShut
  {NoStop}%
\bibitem [{\citenamefont {Duan}\ \emph {et~al.}(2020)\citenamefont {Duan},
  \citenamefont {Sasmal}, \citenamefont {Maple}, \citenamefont {Podlesnyak},
  \citenamefont {Zhu}, \citenamefont {Si},\ and\ \citenamefont {Dai}}]{INS_2}%
  \BibitemOpen
  \bibfield  {author} {\bibinfo {author} {\bibfnamefont {C.}~\bibnamefont
  {Duan}}, \bibinfo {author} {\bibfnamefont {K.}~\bibnamefont {Sasmal}},
  \bibinfo {author} {\bibfnamefont {M.~B.}\ \bibnamefont {Maple}}, \bibinfo
  {author} {\bibfnamefont {A.}~\bibnamefont {Podlesnyak}}, \bibinfo {author}
  {\bibfnamefont {J.-X.}\ \bibnamefont {Zhu}}, \bibinfo {author} {\bibfnamefont
  {Q.}~\bibnamefont {Si}}, \ and\ \bibinfo {author} {\bibfnamefont
  {P.}~\bibnamefont {Dai}},\ }\href {\doibase 10.1103/PhysRevLett.125.237003}
  {\bibfield  {journal} {\bibinfo  {journal} {Phys. Rev. Lett.}\ }\textbf
  {\bibinfo {volume} {125}},\ \bibinfo {pages} {237003} (\bibinfo {year}
  {2020})}\BibitemShut {NoStop}%
\bibitem [{\citenamefont {Butch}\ \emph {et~al.}(2022)\citenamefont {Butch},
  \citenamefont {Ran}, \citenamefont {Saha}, \citenamefont {Neves},
  \citenamefont {Zic}, \citenamefont {Paglione}, \citenamefont {Gladchenko},
  \citenamefont {Ye},\ and\ \citenamefont {Rodriguez-Rivera}}]{Butch_neutron}%
  \BibitemOpen
  \bibfield  {author} {\bibinfo {author} {\bibfnamefont {N.~P.}\ \bibnamefont
  {Butch}}, \bibinfo {author} {\bibfnamefont {S.}~\bibnamefont {Ran}}, \bibinfo
  {author} {\bibfnamefont {S.~R.}\ \bibnamefont {Saha}}, \bibinfo {author}
  {\bibfnamefont {P.~M.}\ \bibnamefont {Neves}}, \bibinfo {author}
  {\bibfnamefont {M.~P.}\ \bibnamefont {Zic}}, \bibinfo {author} {\bibfnamefont
  {J.}~\bibnamefont {Paglione}}, \bibinfo {author} {\bibfnamefont
  {S.}~\bibnamefont {Gladchenko}}, \bibinfo {author} {\bibfnamefont
  {Q.}~\bibnamefont {Ye}}, \ and\ \bibinfo {author} {\bibfnamefont {J.~A.}\
  \bibnamefont {Rodriguez-Rivera}},\ }\href@noop {} {\bibfield  {journal}
  {\bibinfo  {journal} {npj Quantum Materials}\ }\textbf {\bibinfo {volume}
  {7}},\ \bibinfo {pages} {39} (\bibinfo {year} {2022})}\BibitemShut {NoStop}%
\bibitem [{\citenamefont {Knafo}\ \emph {et~al.}(2021)\citenamefont {Knafo},
  \citenamefont {Knebel}, \citenamefont {Steffens}, \citenamefont {Kaneko},
  \citenamefont {Rosuel}, \citenamefont {Brison}, \citenamefont {Flouquet},
  \citenamefont {Aoki}, \citenamefont {Lapertot},\ and\ \citenamefont
  {Raymond}}]{INS_1}%
  \BibitemOpen
  \bibfield  {author} {\bibinfo {author} {\bibfnamefont {W.}~\bibnamefont
  {Knafo}}, \bibinfo {author} {\bibfnamefont {G.}~\bibnamefont {Knebel}},
  \bibinfo {author} {\bibfnamefont {P.}~\bibnamefont {Steffens}}, \bibinfo
  {author} {\bibfnamefont {K.}~\bibnamefont {Kaneko}}, \bibinfo {author}
  {\bibfnamefont {A.}~\bibnamefont {Rosuel}}, \bibinfo {author} {\bibfnamefont
  {J.-P.}\ \bibnamefont {Brison}}, \bibinfo {author} {\bibfnamefont
  {J.}~\bibnamefont {Flouquet}}, \bibinfo {author} {\bibfnamefont
  {D.}~\bibnamefont {Aoki}}, \bibinfo {author} {\bibfnamefont {G.}~\bibnamefont
  {Lapertot}}, \ and\ \bibinfo {author} {\bibfnamefont {S.}~\bibnamefont
  {Raymond}},\ }\href {\doibase 10.1103/PhysRevB.104.L100409} {\bibfield
  {journal} {\bibinfo  {journal} {Phys. Rev. B}\ }\textbf {\bibinfo {volume}
  {104}},\ \bibinfo {pages} {100409} (\bibinfo {year} {2021})}\BibitemShut
  {NoStop}%
\bibitem [{\citenamefont {Wray}\ \emph
  {et~al.}(2015{\natexlab{a}})\citenamefont {Wray}, \citenamefont {Denlinger},
  \citenamefont {Huang}, \citenamefont {He}, \citenamefont {Butch},
  \citenamefont {Maple}, \citenamefont {Hussain},\ and\ \citenamefont
  {Chuang}}]{wray2015}%
  \BibitemOpen
  \bibfield  {author} {\bibinfo {author} {\bibfnamefont {L.~A.}\ \bibnamefont
  {Wray}}, \bibinfo {author} {\bibfnamefont {J.}~\bibnamefont {Denlinger}},
  \bibinfo {author} {\bibfnamefont {S.-W.}\ \bibnamefont {Huang}}, \bibinfo
  {author} {\bibfnamefont {H.}~\bibnamefont {He}}, \bibinfo {author}
  {\bibfnamefont {N.~P.}\ \bibnamefont {Butch}}, \bibinfo {author}
  {\bibfnamefont {M.~B.}\ \bibnamefont {Maple}}, \bibinfo {author}
  {\bibfnamefont {Z.}~\bibnamefont {Hussain}}, \ and\ \bibinfo {author}
  {\bibfnamefont {Y.-D.}\ \bibnamefont {Chuang}},\ }\href@noop {} {\bibfield
  {journal} {\bibinfo  {journal} {Physical Review Letters}\ }\textbf {\bibinfo
  {volume} {114}},\ \bibinfo {pages} {236401} (\bibinfo {year}
  {2015}{\natexlab{a}})}\BibitemShut {NoStop}%
\bibitem [{\citenamefont {Hutanu}\ \emph {et~al.}(2020)\citenamefont {Hutanu},
  \citenamefont {Deng}, \citenamefont {Ran}, \citenamefont {Fuhrman},
  \citenamefont {Thoma},\ and\ \citenamefont {Butch}}]{simu_ref1}%
  \BibitemOpen
  \bibfield  {author} {\bibinfo {author} {\bibfnamefont {V.}~\bibnamefont
  {Hutanu}}, \bibinfo {author} {\bibfnamefont {H.}~\bibnamefont {Deng}},
  \bibinfo {author} {\bibfnamefont {S.}~\bibnamefont {Ran}}, \bibinfo {author}
  {\bibfnamefont {W.~T.}\ \bibnamefont {Fuhrman}}, \bibinfo {author}
  {\bibfnamefont {H.}~\bibnamefont {Thoma}}, \ and\ \bibinfo {author}
  {\bibfnamefont {N.~P.}\ \bibnamefont {Butch}},\ }\href {\doibase
  10.1107/S2052520619016950} {\bibfield  {journal} {\bibinfo  {journal} {Acta
  Crystallographica Section B}\ }\textbf {\bibinfo {volume} {76}},\ \bibinfo
  {pages} {137} (\bibinfo {year} {2020})}\BibitemShut {NoStop}%
\bibitem [{\citenamefont {Rosa}\ \emph {et~al.}(2021)\citenamefont {Rosa},
  \citenamefont {Weiland}, \citenamefont {Fender}, \citenamefont {Scott},
  \citenamefont {Ronning}, \citenamefont {Thompson}, \citenamefont {Bauer},\
  and\ \citenamefont {Thomas}}]{simu_ref2}%
  \BibitemOpen
  \bibfield  {author} {\bibinfo {author} {\bibfnamefont {P.~F.~S.}\
  \bibnamefont {Rosa}}, \bibinfo {author} {\bibfnamefont {A.}~\bibnamefont
  {Weiland}}, \bibinfo {author} {\bibfnamefont {S.~S.}\ \bibnamefont {Fender}},
  \bibinfo {author} {\bibfnamefont {B.~L.}\ \bibnamefont {Scott}}, \bibinfo
  {author} {\bibfnamefont {F.}~\bibnamefont {Ronning}}, \bibinfo {author}
  {\bibfnamefont {J.~D.}\ \bibnamefont {Thompson}}, \bibinfo {author}
  {\bibfnamefont {E.~D.}\ \bibnamefont {Bauer}}, \ and\ \bibinfo {author}
  {\bibfnamefont {S.~M.}\ \bibnamefont {Thomas}},\ }\href {\doibase
  10.48550/ARXIV.2110.06200} {\enquote {\bibinfo {title} {Single-component
  superconducting state in $\mathrm{UTe_2}$ at 2 k},}\ } (\bibinfo {year}
  {2021})\BibitemShut {NoStop}%
\bibitem [{\citenamefont {Kvashnina}\ and\ \citenamefont {{de
  Groot}}(2014)}]{O_edge_importance}%
  \BibitemOpen
  \bibfield  {author} {\bibinfo {author} {\bibfnamefont {K.~O.}\ \bibnamefont
  {Kvashnina}}\ and\ \bibinfo {author} {\bibfnamefont {F.~M.}\ \bibnamefont
  {{de Groot}}},\ }\href {\doibase
  https://doi.org/10.1016/j.elspec.2014.03.012} {\bibfield  {journal} {\bibinfo
   {journal} {Journal of Electron Spectroscopy and Related Phenomena}\ }\textbf
  {\bibinfo {volume} {194}},\ \bibinfo {pages} {88} (\bibinfo {year} {2014})},\
  \bibinfo {note} {core-Level Spectroscopies of Actinides}\BibitemShut
  {NoStop}%
\bibitem [{\citenamefont {Miao}\ \emph {et~al.}(2019)\citenamefont {Miao},
  \citenamefont {Basak}, \citenamefont {Ran}, \citenamefont {Xu}, \citenamefont
  {Kotta}, \citenamefont {He}, \citenamefont {Denlinger}, \citenamefont
  {Chuang} \emph {et~al.}}]{miao_USb2}%
  \BibitemOpen
  \bibfield  {author} {\bibinfo {author} {\bibfnamefont {L.}~\bibnamefont
  {Miao}}, \bibinfo {author} {\bibfnamefont {R.}~\bibnamefont {Basak}},
  \bibinfo {author} {\bibfnamefont {S.}~\bibnamefont {Ran}}, \bibinfo {author}
  {\bibfnamefont {Y.}~\bibnamefont {Xu}}, \bibinfo {author} {\bibfnamefont
  {E.}~\bibnamefont {Kotta}}, \bibinfo {author} {\bibfnamefont
  {H.}~\bibnamefont {He}}, \bibinfo {author} {\bibfnamefont {J.~D.}\
  \bibnamefont {Denlinger}}, \bibinfo {author} {\bibfnamefont {Y.-D.}\
  \bibnamefont {Chuang}},  \emph {et~al.},\ }\href@noop {} {\bibfield
  {journal} {\bibinfo  {journal} {Nature Communications}\ }\textbf {\bibinfo
  {volume} {10}},\ \bibinfo {pages} {644} (\bibinfo {year} {2019})}\BibitemShut
  {NoStop}%
\bibitem [{\citenamefont {Amorese}\ \emph {et~al.}(2020)\citenamefont
  {Amorese}, \citenamefont {Sundermann}, \citenamefont {Leedahl}, \citenamefont
  {Marino}, \citenamefont {Takegami}, \citenamefont {Gretarsson}, \citenamefont
  {Gloskovskii}, \citenamefont {Schlueter}, \citenamefont {Haverkort},
  \citenamefont {Huang}, \citenamefont {Szlawska}, \citenamefont {Kaczorowski},
  \citenamefont {Ran}, \citenamefont {Maple}, \citenamefont {Bauer},
  \citenamefont {Leithe-Jasper}, \citenamefont {Hansmann}, \citenamefont
  {Thalmeier}, \citenamefont {Tjeng},\ and\ \citenamefont
  {Severing}}]{amorese_paper}%
  \BibitemOpen
  \bibfield  {author} {\bibinfo {author} {\bibfnamefont {A.}~\bibnamefont
  {Amorese}}, \bibinfo {author} {\bibfnamefont {M.}~\bibnamefont {Sundermann}},
  \bibinfo {author} {\bibfnamefont {B.}~\bibnamefont {Leedahl}}, \bibinfo
  {author} {\bibfnamefont {A.}~\bibnamefont {Marino}}, \bibinfo {author}
  {\bibfnamefont {D.}~\bibnamefont {Takegami}}, \bibinfo {author}
  {\bibfnamefont {H.}~\bibnamefont {Gretarsson}}, \bibinfo {author}
  {\bibfnamefont {A.}~\bibnamefont {Gloskovskii}}, \bibinfo {author}
  {\bibfnamefont {C.}~\bibnamefont {Schlueter}}, \bibinfo {author}
  {\bibfnamefont {M.~W.}\ \bibnamefont {Haverkort}}, \bibinfo {author}
  {\bibfnamefont {Y.}~\bibnamefont {Huang}}, \bibinfo {author} {\bibfnamefont
  {M.}~\bibnamefont {Szlawska}}, \bibinfo {author} {\bibfnamefont
  {D.}~\bibnamefont {Kaczorowski}}, \bibinfo {author} {\bibfnamefont
  {S.}~\bibnamefont {Ran}}, \bibinfo {author} {\bibfnamefont {M.~B.}\
  \bibnamefont {Maple}}, \bibinfo {author} {\bibfnamefont {E.~D.}\ \bibnamefont
  {Bauer}}, \bibinfo {author} {\bibfnamefont {A.}~\bibnamefont
  {Leithe-Jasper}}, \bibinfo {author} {\bibfnamefont {P.}~\bibnamefont
  {Hansmann}}, \bibinfo {author} {\bibfnamefont {P.}~\bibnamefont {Thalmeier}},
  \bibinfo {author} {\bibfnamefont {L.~H.}\ \bibnamefont {Tjeng}}, \ and\
  \bibinfo {author} {\bibfnamefont {A.}~\bibnamefont {Severing}},\ }\href
  {\doibase 10.1073/pnas.2005701117} {\bibfield  {journal} {\bibinfo  {journal}
  {Proceedings of the National Academy of Sciences}\ }\textbf {\bibinfo
  {volume} {117}},\ \bibinfo {pages} {30220} (\bibinfo {year} {2020})},\
  \Eprint
  {http://arxiv.org/abs/https://www.pnas.org/doi/pdf/10.1073/pnas.2005701117}
  {https://www.pnas.org/doi/pdf/10.1073/pnas.2005701117} \BibitemShut {NoStop}%
\bibitem [{\citenamefont {Okada}\ \emph {et~al.}(1993)\citenamefont {Okada},
  \citenamefont {Kotani}, \citenamefont {Ogasawara}, \citenamefont {Seino},\
  and\ \citenamefont {Thole}}]{Kotani_lifetime}%
  \BibitemOpen
  \bibfield  {author} {\bibinfo {author} {\bibfnamefont {K.}~\bibnamefont
  {Okada}}, \bibinfo {author} {\bibfnamefont {A.}~\bibnamefont {Kotani}},
  \bibinfo {author} {\bibfnamefont {H.}~\bibnamefont {Ogasawara}}, \bibinfo
  {author} {\bibfnamefont {Y.}~\bibnamefont {Seino}}, \ and\ \bibinfo {author}
  {\bibfnamefont {B.~T.}\ \bibnamefont {Thole}},\ }\href@noop {} {\bibfield
  {journal} {\bibinfo  {journal} {Physical Review B}\ }\textbf {\bibinfo
  {volume} {47}},\ \bibinfo {pages} {6203} (\bibinfo {year}
  {1993})}\BibitemShut {NoStop}%
\bibitem [{\citenamefont {Gupta}\ \emph {et~al.}(2011)\citenamefont {Gupta},
  \citenamefont {Bradley}, \citenamefont {Haverkort}, \citenamefont {Seidler},
  \citenamefont {A.},\ and\ \citenamefont {A.}}]{Sawatzky_lifetime}%
  \BibitemOpen
  \bibfield  {author} {\bibinfo {author} {\bibfnamefont {S.~S.}\ \bibnamefont
  {Gupta}}, \bibinfo {author} {\bibfnamefont {J.~A.}\ \bibnamefont {Bradley}},
  \bibinfo {author} {\bibfnamefont {M.~W.}\ \bibnamefont {Haverkort}}, \bibinfo
  {author} {\bibfnamefont {G.~T.}\ \bibnamefont {Seidler}}, \bibinfo {author}
  {\bibfnamefont {T.}~\bibnamefont {A.}}, \ and\ \bibinfo {author}
  {\bibfnamefont {S.~G.}\ \bibnamefont {A.}},\ }\href@noop {} {\bibfield
  {journal} {\bibinfo  {journal} {Physical Review B}\ }\textbf {\bibinfo
  {volume} {84}},\ \bibinfo {pages} {075134} (\bibinfo {year}
  {2011})}\BibitemShut {NoStop}%
\bibitem [{\citenamefont {Wray}\ \emph
  {et~al.}(2015{\natexlab{b}})\citenamefont {Wray}, \citenamefont {Huang},
  \citenamefont {Jarrige}, \citenamefont {Ikeuchi}, \citenamefont {Ishii},
  \citenamefont {Li}, \citenamefont {Qiu}, \citenamefont {Hussain},\ and\
  \citenamefont {Chuang}}]{Frontiers_RIXS_review}%
  \BibitemOpen
  \bibfield  {author} {\bibinfo {author} {\bibfnamefont {L.~A.}\ \bibnamefont
  {Wray}}, \bibinfo {author} {\bibfnamefont {S.-W.}\ \bibnamefont {Huang}},
  \bibinfo {author} {\bibfnamefont {I.}~\bibnamefont {Jarrige}}, \bibinfo
  {author} {\bibfnamefont {K.}~\bibnamefont {Ikeuchi}}, \bibinfo {author}
  {\bibfnamefont {K.}~\bibnamefont {Ishii}}, \bibinfo {author} {\bibfnamefont
  {J.}~\bibnamefont {Li}}, \bibinfo {author} {\bibfnamefont {Z.~Q.}\
  \bibnamefont {Qiu}}, \bibinfo {author} {\bibfnamefont {Z.}~\bibnamefont
  {Hussain}}, \ and\ \bibinfo {author} {\bibfnamefont {Y.-D.}\ \bibnamefont
  {Chuang}},\ }\href@noop {} {\bibfield  {journal} {\bibinfo  {journal} {Front.
  Phys.}\ }\textbf {\bibinfo {volume} {3}},\ \bibinfo {pages} {32} (\bibinfo
  {year} {2015}{\natexlab{b}})}\BibitemShut {NoStop}%
\bibitem [{\citenamefont {Carra}\ \emph {et~al.}(1993)\citenamefont {Carra},
  \citenamefont {K{\"o}nig}, \citenamefont {Thole},\ and\ \citenamefont
  {Altarelli}}]{linear_dichroism_quadrupole}%
  \BibitemOpen
  \bibfield  {author} {\bibinfo {author} {\bibfnamefont {P.}~\bibnamefont
  {Carra}}, \bibinfo {author} {\bibfnamefont {H.}~\bibnamefont {K{\"o}nig}},
  \bibinfo {author} {\bibfnamefont {B.}~\bibnamefont {Thole}}, \ and\ \bibinfo
  {author} {\bibfnamefont {M.}~\bibnamefont {Altarelli}},\ }\href@noop {}
  {\bibfield  {journal} {\bibinfo  {journal} {Physica B: Condensed Matter}\
  }\textbf {\bibinfo {volume} {192}},\ \bibinfo {pages} {182} (\bibinfo {year}
  {1993})}\BibitemShut {NoStop}%
\bibitem [{\citenamefont {Ament}\ \emph {et~al.}(2011)\citenamefont {Ament},
  \citenamefont {van Veenendaal}, \citenamefont {Devereaux}, \citenamefont
  {Hill},\ and\ \citenamefont {van~den Brink}}]{rixs_review}%
  \BibitemOpen
  \bibfield  {author} {\bibinfo {author} {\bibfnamefont {L.~J.~P.}\
  \bibnamefont {Ament}}, \bibinfo {author} {\bibfnamefont {M.}~\bibnamefont
  {van Veenendaal}}, \bibinfo {author} {\bibfnamefont {T.~P.}\ \bibnamefont
  {Devereaux}}, \bibinfo {author} {\bibfnamefont {J.~P.}\ \bibnamefont {Hill}},
  \ and\ \bibinfo {author} {\bibfnamefont {J.}~\bibnamefont {van~den Brink}},\
  }\href {\doibase 10.1103/RevModPhys.83.705} {\bibfield  {journal} {\bibinfo
  {journal} {Rev. Mod. Phys.}\ }\textbf {\bibinfo {volume} {83}},\ \bibinfo
  {pages} {705} (\bibinfo {year} {2011})}\BibitemShut {NoStop}%
\bibitem [{SI()}]{SI}%
  \BibitemOpen
  \href@noop {} {}\bibinfo {note} {See Supplemental Material at [URL will be
  inserted by publisher] for additional discussion of fitting details and
  excitation symmetries.}\BibitemShut {Stop}%
\bibitem [{\citenamefont {Mekonen}\ \emph {et~al.}(2022)\citenamefont
  {Mekonen}, \citenamefont {Kang}, \citenamefont {Chaudhuri}, \citenamefont
  {Barbalas}, \citenamefont {Ran}, \citenamefont {Kotliar}, \citenamefont
  {Butch},\ and\ \citenamefont {Armitage}}]{Armitage_optical}%
  \BibitemOpen
  \bibfield  {author} {\bibinfo {author} {\bibfnamefont {S.~M.}\ \bibnamefont
  {Mekonen}}, \bibinfo {author} {\bibfnamefont {C.-J.}\ \bibnamefont {Kang}},
  \bibinfo {author} {\bibfnamefont {D.}~\bibnamefont {Chaudhuri}}, \bibinfo
  {author} {\bibfnamefont {D.}~\bibnamefont {Barbalas}}, \bibinfo {author}
  {\bibfnamefont {S.}~\bibnamefont {Ran}}, \bibinfo {author} {\bibfnamefont
  {G.}~\bibnamefont {Kotliar}}, \bibinfo {author} {\bibfnamefont {N.~P.}\
  \bibnamefont {Butch}}, \ and\ \bibinfo {author} {\bibfnamefont {N.~P.}\
  \bibnamefont {Armitage}},\ }\href@noop {} {\bibfield  {journal} {\bibinfo
  {journal} {Physical Review B}\ }\textbf {\bibinfo {volume} {106}},\ \bibinfo
  {pages} {085125} (\bibinfo {year} {2022})}\BibitemShut {NoStop}%
\bibitem [{\citenamefont {Shick}\ \emph {et~al.}(2021)\citenamefont {Shick},
  \citenamefont {Fujimori},\ and\ \citenamefont {Pickett}}]{fujimori_paper}%
  \BibitemOpen
  \bibfield  {author} {\bibinfo {author} {\bibfnamefont {A.~B.}\ \bibnamefont
  {Shick}}, \bibinfo {author} {\bibfnamefont {S.-i.}\ \bibnamefont {Fujimori}},
  \ and\ \bibinfo {author} {\bibfnamefont {W.~E.}\ \bibnamefont {Pickett}},\
  }\href {\doibase 10.1103/PhysRevB.103.125136} {\bibfield  {journal} {\bibinfo
   {journal} {Phys. Rev. B}\ }\textbf {\bibinfo {volume} {103}},\ \bibinfo
  {pages} {125136} (\bibinfo {year} {2021})}\BibitemShut {NoStop}%
\bibitem [{\citenamefont {van Elp}\ \emph {et~al.}(1992)\citenamefont {van
  Elp}, \citenamefont {Eskes}, \citenamefont {Kuiper},\ and\ \citenamefont
  {Sawatzky}}]{Elp_NiO}%
  \BibitemOpen
  \bibfield  {author} {\bibinfo {author} {\bibfnamefont {J.}~\bibnamefont {van
  Elp}}, \bibinfo {author} {\bibfnamefont {H.}~\bibnamefont {Eskes}}, \bibinfo
  {author} {\bibfnamefont {P.}~\bibnamefont {Kuiper}}, \ and\ \bibinfo {author}
  {\bibfnamefont {G.~A.}\ \bibnamefont {Sawatzky}},\ }\href@noop {} {\bibfield
  {journal} {\bibinfo  {journal} {Physical Review B}\ }\textbf {\bibinfo
  {volume} {45}},\ \bibinfo {pages} {1612} (\bibinfo {year}
  {1992})}\BibitemShut {NoStop}%
\bibitem [{\citenamefont {Haverkort}\ \emph {et~al.}(2005)\citenamefont
  {Haverkort}, \citenamefont {Hu}, \citenamefont {Tanaka}, \citenamefont
  {Reichelt}, \citenamefont {Streltsov}, \citenamefont {Korotin}, \citenamefont
  {Anisimov}, \citenamefont {Hsieh}, \citenamefont {Lin}, \citenamefont {Chen},
  \citenamefont {Khomskii},\ and\ \citenamefont {Tjeng}}]{VO2_paper}%
  \BibitemOpen
  \bibfield  {author} {\bibinfo {author} {\bibfnamefont {M.~W.}\ \bibnamefont
  {Haverkort}}, \bibinfo {author} {\bibfnamefont {Z.}~\bibnamefont {Hu}},
  \bibinfo {author} {\bibfnamefont {A.}~\bibnamefont {Tanaka}}, \bibinfo
  {author} {\bibfnamefont {W.}~\bibnamefont {Reichelt}}, \bibinfo {author}
  {\bibfnamefont {S.~V.}\ \bibnamefont {Streltsov}}, \bibinfo {author}
  {\bibfnamefont {M.~A.}\ \bibnamefont {Korotin}}, \bibinfo {author}
  {\bibfnamefont {V.~I.}\ \bibnamefont {Anisimov}}, \bibinfo {author}
  {\bibfnamefont {H.~H.}\ \bibnamefont {Hsieh}}, \bibinfo {author}
  {\bibfnamefont {H.-J.}\ \bibnamefont {Lin}}, \bibinfo {author} {\bibfnamefont
  {C.~T.}\ \bibnamefont {Chen}}, \bibinfo {author} {\bibfnamefont {D.~I.}\
  \bibnamefont {Khomskii}}, \ and\ \bibinfo {author} {\bibfnamefont {L.~H.}\
  \bibnamefont {Tjeng}},\ }\href {\doibase 10.1103/PhysRevLett.95.196404}
  {\bibfield  {journal} {\bibinfo  {journal} {Phys. Rev. Lett.}\ }\textbf
  {\bibinfo {volume} {95}},\ \bibinfo {pages} {196404} (\bibinfo {year}
  {2005})}\BibitemShut {NoStop}%
\end{thebibliography}%
\end{document}